\documentclass[]{elsart}
\usepackage[latin1]{inputenc}
\usepackage{times}
\usepackage{latexsym}
\usepackage{epsf}
\begin{document}
\begin{frontmatter}
\title{\Large Influence of boundary conditions on\\ level statistics and 
eigenstates \\ at the  metal insulator transition}
\author{L. Schweitzer and H. Potempa}
\address{Physikalisch-Technische Bundesanstalt, Bundesallee 100, 
38116 Braunschweig, Germany}

\begin{abstract}
We investigate the influence of the boundary conditions on the scale
invariant critical level statistics at the metal insulator transition
of disordered three-dimensional orthogonal and two-dimensional unitary 
and symplectic tight-binding models.
The distribution of the spacings between consecutive eigenvalues is
calculated numerically and shown to be different for periodic and 
Dirichlet boundary conditions whereas the critical disorder remains
unchanged. The peculiar correlations of the corresponding critical 
eigenstates leading to anomalous diffusion seem not to be affected by 
the change of the boundary conditions.
\end{abstract}
\end{frontmatter}
\section{Introduction and Model}
The statistics of energy eigenvalues has proven to be a powerful tool 
to describe the localization-delocalization transition (LDT) in disordered 
electronic systems. For three-dimensional models, numerical work
\cite{Sea93,HS93a,HS94a,ZK95,BSZK96,ZK97} 
and analytical investigations \cite{AKL94,KM94,AKL95,CLS96,CKL96}
have revealed a variety of new features and important relations 
that constitute the peculiar dynamics at the metal insulator transition.
Also two-dimensional systems that exhibit a LDT were investigated numerically,
e.g., electrons in a strong perpendicular magnetic field which show the 
quantum Hall effect (QHE) \cite{OO95,FAB95,BS97}
or in the presence of spin-orbit interaction leading to a symplectic symmetry
\cite{SZ95,Eva95,SZ97}. 

It has been suggested recently that the eigenvalue statistics directly at the 
critical point depend on the boundary conditions \cite{BMP98} and also on the 
shape of the sample \cite{PS98} even in the limit of infinite system size.
The so called critical statistics are scale independent and distinct from both
the random matrix theory result which is appropriate for disordered systems 
with conducting electrons (extended states) and the Poisson law that describes 
the statistics of uncorrelated eigenvalues typically found for insulating 
(localized) states. The probability of eigenvalues being close in energy
is drastically reduced for extended states (level repulsion) while it is
maximal for localized states due to the negligible overlap of the 
corresponding eigenstates.

The reported dependence of the critical level statistics on 
the boundary conditions \cite{BMP98} and on the 
sample shape \cite{PS98} seems at first glance to be somewhat counter 
intuitive because in general for macroscopic samples one expects physical 
observables not to be sensitive to boundary effects. 
However, for mesoscopic systems it is known that the preserved phase coherence
gives rise to various macroscopic quantum effects originating from, e.g., an 
applied Aharonov-Bohm flux which can be completely incorporated into the 
boundary conditions. 
It remains, however, to be seen whether or not the observed sensitivity of the
level statistics with respect to changes of the boundary conditions will have
an effect on measurable physical quantities.
A related question is the possible influence of the boundary conditions on 
the critical wave functions which were shown to be multifractal objects 
\cite{Weg80a,Aok83,Aok86,CP86,Eva90a,SG91,HS92,PJ91,HKS92,Sch95}. 
The strong amplitude fluctuations of the eigenstates are responsible for 
anomalous diffusion which can be described by correlation functions 
characterized by an exponent $\eta$ \cite{CD88,HS94,BHS96}.

In this paper, we address the questions whether the influence on the boundary 
conditions is also present in 2d critical systems and if the 
eigenstates of the 3d Anderson model are affected too. 
Therefore, we present results of a numerical investigation of the
critical eigenvalue statistics and the correlations of the corresponding 
eigenvectors. We consider standard tight binding Hamiltonians with diagonal 
disorder on a simple cubic lattice for 3d (Anderson model) and on square 
lattices for the 2d QHE \cite{SKM84a} and the 2d symplectic model 
\cite{And89}. 
The disorder potentials $\{\varepsilon_n \}$ are independent random numbers
evenly distributed around zero. The width of this box distribution 
determines the disorder strength $W$.
The Hamiltonian for the 3d orthogonal case (preserved time reversal symmetry) 
is given by 
\begin{equation}
H=\sum_n \varepsilon_n c_n^\dagger c_n^{} + 
\sum_{<m\ne n>}V_{mn}c_m^\dagger c_n^{},
\label{hamilton}
\end{equation}
where $c_n^\dagger, c_n^{}$ are the creation and annihilation operators, 
respectively, and $\{m, n\}$ denote the sites on the cubic lattice with 
lattice constant $a$.
The transfer $V_{mn}\equiv V=1$ is restricted to nearest neighbors only and
defines the unit of energy.  
For the 2d unitary system (broken time reversal symmetry) the Hamilton 
operator is the same (Eq.~\ref{hamilton}), except that $\{m, n\}$ now 
represent the sites of a square lattice. Here, the magnetic field $B$ is
incorporated into the complex phase factors of the transfer terms, 
$V_{mn}=V\exp(i2\pi e/h \int_{\mathbf{r}_m}^{\mathbf{r}_n} 
\mathbf{A}(\mathbf{r})\,\mathrm{d}\mathbf{r})$, and the Landau gauge is chosen 
for the vector potential $\mathbf{A}=(0,-B\mathbf{x},0)$. 
The 2d symplectic model (broken spin-rotational invariance) is described by
\cite{And89}
\begin{equation}
H = \sum_{m,\sigma } \varepsilon^{} _{m} c^{\dagger}_{m,\sigma }
c^{}_{m,\sigma} + \sum_{<m\ne n>, \sigma \sigma '} 
V(m,\sigma ; n,\sigma ')\,c^{\dagger}_{m,\sigma } c^{}_{n,\sigma '}.
\end{equation}
The spin-orbit interaction strength $S$ is defined as the ratio 
$S=V_2/(V_1^2+V_2^2)^{1/2}$, where $V_1$ and $V_2$ are matrix elements of 
the $2\times 2$ complex transition matrices $V(m,\sigma ; n,\sigma ')$ 
\cite{And89}, which depend on the transfer-direction and on spin $\sigma$. 
In the following we choose the maximal value $S=0.5$ and define the unit of 
energy by $V\equiv(V_1^2+V_2^2)^{1/2}=1$. The eigenvalues are calculated
using a Lanczos algorithm and care is taken to properly unfold the spectrum
in order to distinguish local fluctuations of the eigenvalues from possible
global changes in the density of states. 

\begin{figure}\centering
\epsfxsize12cm\epsfbox{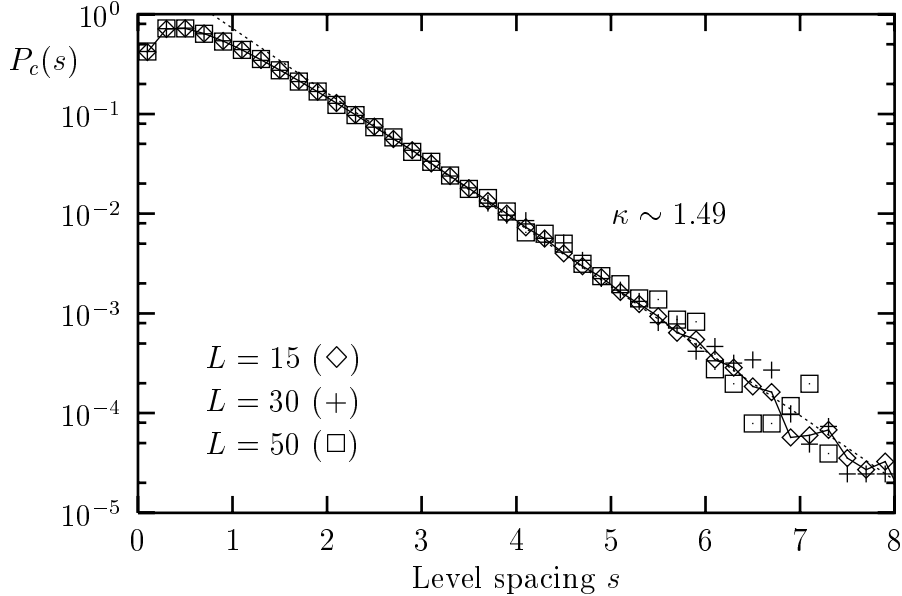}
\caption[]{Critical level spacing distribution, $P_c(s)$, for a 3d 
orthogonal system with Dirichlet boundary conditions (DBC). \label{fig1}}
\end{figure}

\section{Results and Discussions}
The nearest neighbor level spacing distribution, $P_c(s)$, of the 3d 
orthogonal system (Anderson model) is shown in Fig.~\ref{fig1} for Dirichlet 
boundary conditions (DBC).  As usual, the spacings $s=|E_{i+1}-E_i|/\Delta$ 
of successive eigenenergies $\{E_i\}$ 
are divided by the mean level spacing $\Delta$. The curves are obtained at 
the critical disorder $W_c=16.4\,V$ for different system sizes $L/a=15$, 30, 
and 50, which is more than a factor of 2 larger than in \cite{BMP98}, but still
no size dependence can be observed. The corresponding values of the second
moment, $J_c\equiv\langle s^2\rangle=\int_0^\infty s^2 P_c(s)\,ds$, are   
calculated to be $J_c(L/a=15)=1.608$, $J_c(L/a=30)=1.617$, 
and $J_c(L/a=50)=1.616$, 
which are in accord with those of Ref.~\cite{BMP98}. 
While for small $s$, $P_c(s)\sim s$, as expected from RMT, 
the large-$s$ behavior is well approximated by a simple 
exponential decay, $P_c(s) \sim \exp(-\kappa s)$, with 
$\kappa_{\rm DBC}\approx 1.49$ which is significantly 
smaller than the result for periodic boundary conditions (PBC), 
$\kappa_{\rm DBC} \approx 1.9$ \cite{ZK95,BSZK96}.

\begin{figure}\centering
\epsfxsize12cm\epsfbox{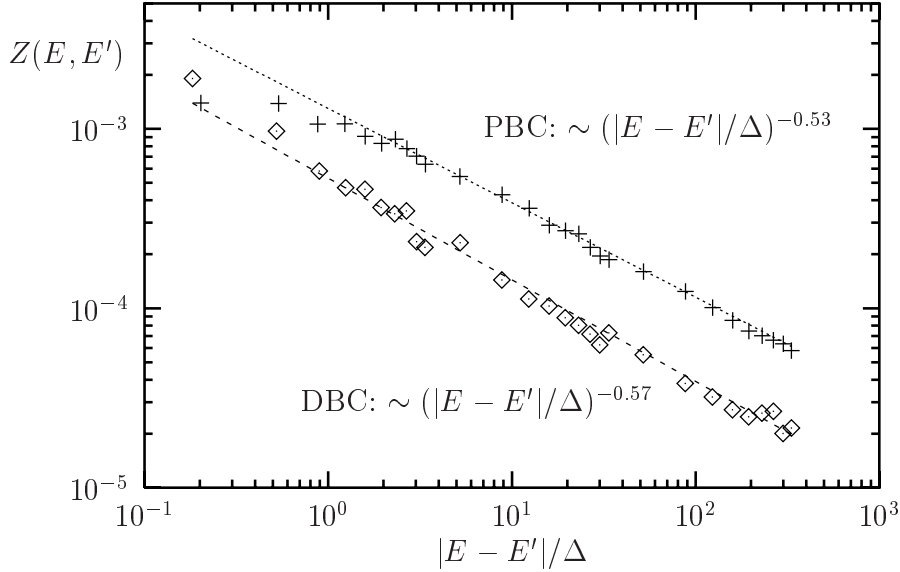}
\caption[]{Critical eigenfunction correlations for a 3d system of size 
$L=40\,a$ 
with Dirichlet (DBC) and periodic boundary conditions (PBC).\label{corr}}
\end{figure}

A correlation function of the corresponding critical eigenfunctions 
$\psi_E(x)$ is shown in Fig.~\ref{corr} for a system of linear size 
$L=40\,a$ and different boundary conditions. As in Ref.~\cite{BHS96} 
we compute the function
\begin{equation}
Z(E,E')= \int_0^\infty |\psi_E(x)|^2 |\psi_{E'}(x)|^2 d^dx \sim 
\left(\frac{|E-E'|}{\Delta}\right)^{-\mu}
\end{equation}
from which the exponent $\mu=\eta/d$ can be extracted. $\eta$ is related to 
the fractal correlation dimension $D(2)=d-\eta$ \cite{PJ91}. 
In the logarithmic plot the power law behavior 
$Z(E,E')\sim (|E-E'|/\Delta)^{-\mu}$ becomes apparent. 
We find two exponents $\mu_{PBC}=0.53\pm 0.1$ and $\mu_{DBC}=0.57\pm 0.1$ 
which are indistinguishable within the uncertainty of our data. 

\begin{figure}\centering
\epsfxsize12cm\epsfbox{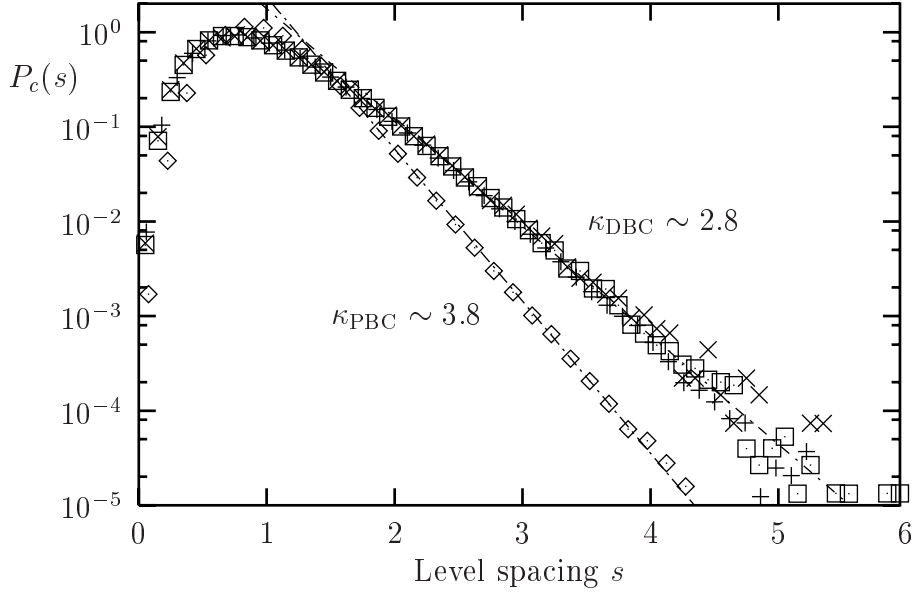}
\caption[]{Critical level spacing distribution of 2d symplectic systems 
with PBC of size $L=20\,a$ ($\Diamond$), and with DBC of size $L=40\,a$ ($+$),
$L=80\,a$ ($\Box$), and $L=120\,a$ ($\times$). The large-s behavior is well 
fitted by $P_c(s)\sim\exp(-\kappa s)$.\label{fig3}}
\end{figure}

A similar behavior as in 3d is observed for the level statistics of the
two-dimensional system with spin-orbit interaction that shows also a complete 
metal insulator transition at a critical disorder $W_c\simeq 6.0\,V$ 
\cite{SZ97}. Within numerical uncertainty the critical $P_c(s)$ for DBC is 
scale independent and distinct from the one with PBC (see Fig.~\ref{fig3}). 
This manifests itself in the different decay constants $\kappa_{\rm PBC}
= 3.8\pm 0.2$ and $\kappa_{\rm DBC} = 2.8\pm 0.2$. 
The values for the second moments of the critical symplectic distributions 
are $J_c^{PBC}=1.142$ and $J_c^{DBC}=1.254$. 
The attempts to look for another critical disorder at which $P_c(s)$ for
DBC coincides with the level spacing distribution for PBC at 
$W_c\simeq 6.0\,V$, as suggested in Ref.~\cite{EK98} for the QHE system, 
ended without any result. 

The situation for the 2d QHE-system is similar, however, due to the
incomplete metal insulator transition the case is more complicated.
In a QHE-system all states are localized with a localization length diverging 
at singular energies $E_n$. At these critical points the eigenstates are 
multifractal \cite{Aok86,HS92,PJ91,HKS92} and a normal extended phase is 
absent. The application of Dirichlet boundary conditions introduces edge 
states that extend along the sample boundaries which seems to cause a certain
shift of the critical energies $E_n$ \cite{EK98}. 
The question remains, whether the small differences that we found between the 
two distributions will show up also in the limit of infinite system size.    

In conclusion, we have shown that the scale independent critical energy level 
statistics are influenced by the boundary conditions in 2d and 3d systems. 
For the 3d Anderson model, the correlations of the corresponding eigenstates 
seem not to be affected by a change of the boundary conditions, at least 
within the uncertainty of our calculations.


\end{document}